\documentclass[aps,pra, twocolumn, amsmath, amssymb, showpacs, superscriptaddress]{revtex4}
\usepackage{graphicx}
\usepackage{bm}

\usepackage{xcolor}

\usepackage{ulem}

\begin{document}
\title{Probe field ellipticity-induced shift in an atomic clock}
\author{V. I. Yudin}
\email{viyudin@mail.ru}
\affiliation{Institute of Laser Physics SB RAS, pr. Akademika Lavrent'eva 15 B, Novosibirsk, 630090, Russia}
\affiliation{Novosibirsk State University, ul. Pirogova 1, Novosibirsk, 630090, Russia}
\affiliation{Novosibirsk State Technical University, pr. Karla Marksa 20, Novosibirsk, 630073, Russia}
\author{A. V. Taichenachev}
\affiliation{Institute of Laser Physics SB RAS, pr. Akademika Lavrent'eva 15 B, Novosibirsk, 630090, Russia}
\affiliation{Novosibirsk State University, ul. Pirogova 1, Novosibirsk, 630090, Russia}
\author{O. N. Prudnikov}
\affiliation{Institute of Laser Physics SB RAS, pr. Akademika Lavrent'eva 15 B, Novosibirsk, 630090, Russia}
\affiliation{Novosibirsk State University, ul. Pirogova 1, Novosibirsk, 630090, Russia}
\author{M.~Yu.~Basalaev}
\affiliation{Institute of Laser Physics SB RAS, pr. Akademika Lavrent'eva 15 B, Novosibirsk, 630090, Russia}
\affiliation{Novosibirsk State University, ul. Pirogova 1, Novosibirsk, 630090, Russia}
\affiliation{Novosibirsk State Technical University, pr. Karla Marksa 20, Novosibirsk, 630073, Russia}
\author{V.~G.~Pal'chikov}
\affiliation{All-Russian Research Institute of Physical and Radio Engineering Measurements, 141570 Mendeleevo, Moscow region, Russia}
\affiliation{National Nuclear Research University MEPhI, Kashirskoe sh. 31, 115409 Moscow, Russia}
\author{M.~von~Boehn}
\affiliation{Physikalisch-Technische Bundesanstalt, Bundesallee 100, D-38116 Braunschweig, Germany}
\author{T.~E.~Mehlst\"aubler}
\affiliation{Physikalisch-Technische Bundesanstalt, Bundesallee 100, D-38116 Braunschweig, Germany}
\affiliation{Institut f\"{u}r Quantenoptik, Leibniz Universit\"{a}t Hannover, Welfengarten 1, 30167 Hannover, Germany}
\author{S. N. Bagayev}
\affiliation{Institute of Laser Physics SB RAS, pr. Akademika Lavrent'eva 15 B, Novosibirsk, 630090, Russia}
\affiliation{Novosibirsk State University, ul. Pirogova 1, Novosibirsk, 630090, Russia}

\date{\today}

\begin{abstract}
We investigate the probe field induced shift for atomic lattice-based and ion-trap clocks, which can be considered as a near resonant ac-Stark shift, connected to the Zeeman structure of atomic levels and their splitting in a dc magnetic field. This shift arises from possible residual ellipticity in the polarization of the probe field and uncertainty in the magnetic field orientation.  Such a shift can have an arbitrary sign and, for some experimental conditions, can reach the fractional value of the order of 10$^{-18}$-10$^{-19}$, i.e., it is not negligible. Thus, it should be taken into account in the uncertainty budgets for the modern ultra-precise atomic clocks. In addition, it is shown that when using hyper-Ramsey spectroscopy, this shift can be reduced to a level much lower than $10^{-19}$.

\end{abstract}


\maketitle

\section{Introduction}
Ultra-precise atomic clocks are the forefront of modern quantum sensors and tests of the standard model \cite{Ludlow_2015,
Safronova_2018,Tanja_2018,Riehle_2017,Delva_2017}. At present, some laboratories have demonstrated systematic uncertainties and long-term instabilities at a fractional level of $10^{-18}$ both for devices with neutral atoms trapped in an optical lattice at the magic wavelength \cite{Katori_2015,Nicholson_2015,Ye_2017,Schioppo_2017,McGrew_2018,Ye_2019,Bothwell_2019}, and for clocks using trapped ions \cite{Huntemann_2016,Brewer_2019}. There is a recent trend to push fractional uncertainties to the level of $10^{-19}$  \cite{Brewer_2019,Marti_2018,Keller_2019}. However, such an extraordinary high metrological precision requires a very thorough study of all possible frequency shifts that can exceed (at least in principle) the value of $10^{-19}$.

\begin{figure}[t]
\centerline{\scalebox{0.35}{\includegraphics{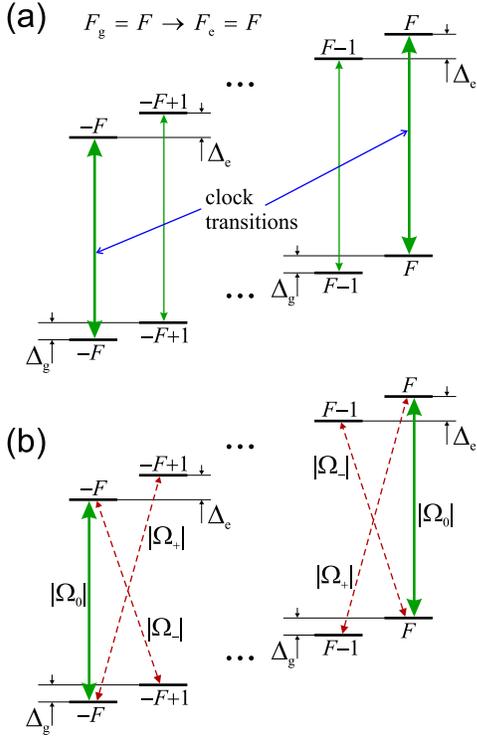}}}\caption{
Scheme of light-induced transitions over Zeeman structure for an optical transition $F_{\rm g}=F\rightarrow F_{\rm e}=F$:\\
(a) the ideal case of linear polarization of the probe field ${\bf E}$ directed along the magnetic field ${\bf B}$ (i.e., ${\bf E}||{\bf B}$);\\
(b) the general case of elliptical polarization of the probe field ${\bf E}$ under arbitrary orientation of the magnetic field ${\bf B}$, where the dotted red lines mark the light-induced transitions leading to ac Stark shifts (\ref{D_FF_ac}) for the clock transitions $|m_{\rm g}= -F\rangle\rightarrow |m_{\rm e}=-F\rangle$ and $|m_{\rm g}=+F\rangle\rightarrow |m_{\rm e}=+F\rangle$.} \label{Fig1}
\end{figure}

In this paper, we investigate the ac Stark shift due to some small residual ellipticity of the probe field, which has not previously been widely discussed in the scientific literature. We consider in detail the case of clock transitions $F_{\rm g}=F\rightarrow F_{\rm e}=F$ (where $F_{\rm g}$ and $F_{\rm e}$ are the angular momenta of the energy levels in the ground and excited states, respectively). In particular, such a variant takes place for strongly forbidden electronic transitions $^1\mathrm{S}_0\rightarrow{}^3$P$_0$ for odd isotopes of alkaline earth neutral atoms (e.g., $^{87}$Sr, $^{171}$Yb) and for some ions (e.g., with two remaining valence electrons, such as $^{27}$Al$^+$, $^{115}$In$^+$). It is shown that this shift can reach a fractional level of the order of $10^{-18}$  for some experimental conditions and therefore, needs to be taken into account in the uncertainty budget for modern ultra-precise atomic clocks. In addition, we propose a method for the radical suppression of any ac Stark shift by using hyper-Ramsey spectroscopy Ref.~\cite{yudin2010}.

\section{\label{Model}Theoretical model}
Let us consider an electric-dipole ($E1$) clock transition $F_{\rm g}=F\rightarrow F_{\rm e}=F$ with an unperturbed frequency $\omega^{}_0$, where $F_{\rm g}$ and $F_{\rm e}$ are the angular momenta of the energy levels in the ground and excited states, respectively. As noted above, this type of transition takes place in atomic clocks based on the strongly forbidden $^1$S$_0\rightarrow{}^3$P$_0$ intercombination transition in alkaline earth (and similar) atoms (Mg, Ca, Sr, Yb, Hg), as well as for some ions (Al$^+$, In$^+$). For these elements, the value of $F$ is determined by the nuclear spin, which is a half-integer (odd isotopes). Under the presence of an external magnetic field ${\bf B}$, the Zeeman splitting of the levels in magnetic sublevels $|m_{\rm g}\rangle$ and $|m_{\rm e}\rangle$ occurs due to the nonzero magnetic moment, where $ |m_{\rm g,e}|\leq F$ (see Fig.~\ref{Fig1}). To eliminate the linear magnetic sensitivity in atomic clocks, the following measurement procedure is usually used. A probe light field is chosen with a linear polarization vector ${\bf E}$ parallel to the dc magnetic field, ${\bf E}||{\bf B}$. For this case, possible light-induced transitions between Zeeman sublevels are shown in Fig.~\ref{Fig1}(a). Next, by pumping atoms in turn to the extreme Zeeman sublevels of the ground state $|m_{\rm g}=\pm F\rangle$, two frequencies $\omega^{}_{-F-F}$ and $\omega^{}_{+F+F}$ are successively measured on the optical clock transitions $|m_{\rm g}=-F\rangle\rightarrow |m_{\rm e}=-F\rangle$ and $|m_{\rm g}= +F\rangle\rightarrow |m_{\rm e}=+F\rangle$, respectively [see Fig.~\ref{Fig1}(a)]. Each of these frequencies separately experiences a linear Zeeman shift with respect to the unperturbed frequency $\omega^{}_0$:
\begin{eqnarray}\label{w_FF}
\omega^{}_{-F-F}&=&\omega^{}_0-(\Delta^{}_{\rm e}-\Delta^{}_{\rm g})F\,,\nonumber\\
\omega^{}_{+F+F}&=&\omega^{}_0+(\Delta^{}_{\rm e}-\Delta^{}_{\rm g})F\,,
\end{eqnarray}
where
\begin{equation}\label{D}
\Delta^{}_{\rm g}=g^{}_{\rm g}\mu^{}_B|{\bf B}|/\hbar\,,\quad \Delta^{}_{\rm e}=g^{}_{\rm e}\mu^{}_B|{\bf B}|/\hbar\,,
\end{equation}
are the Zeeman splittings in the ground and excited states, respectively [see Fig.~\ref{Fig1}(a)]; $\mu^{}_B$ is the Bohr magneton; and $g^{}_{\rm g}$ and $g^{}_{\rm e}$ are the $g$-factors in the ground and excited states, respectively. However, for the superposition frequency
\begin{equation}\label{w_clock}
  \omega_{\rm clock}=\frac{\omega^{}_{-F-F}+\omega^{}_{+F+F}}{2}=\omega^{}_0\,,
\end{equation}
considered as a clock frequency, there is no linear sensitivity to a magnetic field. The residual quadratic sensitivity to the magnetic field, $\propto |{\bf B}|^2$, is due to the non-resonant interaction between different energy levels induced by the magnetic-dipole interaction operator $-({\hat{\bf \mu}{\bf B}})$, where $\hat{\bf \mu}$ is the operator of the magnetic moment of an atom (ion). For example, for the above mentioned elements, this occurs as a result of the magnetic-dipole interaction between the $^3$P$_0$ and $^3$P$_1$ fine structure states.

\begin{figure}[t]
\centerline{\scalebox{0.4}{\includegraphics{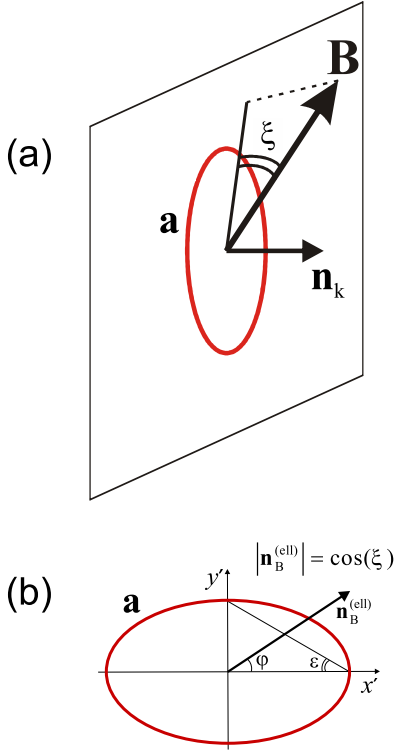}}}\caption{
(a) the general case of an arbitrary orientation of dc magnetic field ${\bf B}$ with respect to the polarization ellipse ${\bf a}$, where $\xi$ is the angle between the magnetic field vector and the plane of the polarization ellipse; \\
(b) the geometric parametrization of the elliptical polarization ${\bf a}$ according to Eq.~(\ref{eps}); ${\bf n}^{\rm (ell)}_{\rm B}$ is the projection of the unit vector ${\bf n}^{}_{\rm B}={\bf B}/|{\bf B}|$ onto the plane of the polarization ellipse, and $\varphi$ is the angle between ${\bf n }^{\rm (ell)}_{\rm B}$ and the major axis of the polarization ellipse.}\label{Fig2}
\end{figure}

However, the above ideal picture of interaction [see Fig.~\ref{Fig1}(a)] in reality can be violated due to the presence of an uncontrolled ellipticity of the probe field, and also due to an uncertainty in the direction of the magnetic field vector. In the general case of arbitrary orientation of the magnetic field ${\bf B}$ relative to the polarization ellipse [see Fig.~\ref{Fig2}(a)], the electric vector of the probe resonant field (with frequency $\omega$) has the following form
\begin{eqnarray}\label{E}
&&{\bf E}(t)={\rm Re}\{Ee^{-i\omega t}{\bf a}\}=(Ee^{-i\omega t}{\bf a}+c.c.)/2\,,\nonumber \\
&&{\bf a}=\sum^{}_{q=0,\pm 1}a^{(q)}{\bf e}^{}_{q}\,,\quad \sum^{}_{q=0,\pm 1}|a^{(q)}|^2=1\,,
\end{eqnarray}
where $E$ is the scalar amplitude; $a^{(q)}$ are the contravariant components of the unit complex vector of polarization ${\bf a}$ (i.e., $|{\bf a}|=1$) in the cyclic basis ${\bf e}^{}_0={\bf e}^{}_z$, ${\bf e}^{}_{\pm 1}=\mp ({\bf e}^{}_x\pm i{\bf e}^{}_y)/\sqrt{2}$ (where ${\bf e}^{}_x$, ${\bf e}^{}_y$, ${\bf e}^{}_z$ are unit basis vectors of the Cartesian coordinate system, for which the quantization axis $Oz$ is directed along the magnetic vector ${\bf B}$). In the case of the electric-dipole interaction operator $-(\hat{{\bf d}}{\bf E})$, the general picture of light-induced transitions near the extreme Zeeman sublevels is shown in Fig.~\ref{Fig1}(b), where the following expressions hold for the corresponding Rabi frequencies (their absolute values):
\begin{equation}\label{W}
|\Omega^{}_{0}|=\frac{|d^{}_{\rm eg}Ea^{(0)}|\sqrt{F}}{\hbar\sqrt{F+1}}\,,\;\; |\Omega^{}_{\pm}|=\frac{|d^{}_{\rm eg}Ea^{(\pm 1)}|}{\hbar\sqrt{F+1}}\,,
\end{equation}
where $d^{}_{\rm eg}=\langle F^{}_{\rm e} ||\hat{{\bf d}}||F^{}_{\rm g}\rangle$ is the reduced matrix element of the dipole moment operator $\hat{{\bf d}}$ for the clock transition $F_{\rm g}=F\rightarrow F_{\rm e}=F$.

Let us consider the effect of transitions caused by the presence of two circular components $a^{(\pm 1)}$ of the polarization vector ${\bf a}$ [see Eq.~(\ref{E})], which correspond to the Rabi frequencies $|\Omega^{}_{\pm}|$ [see Fig.~\ref{Fig1}(b)]. Since, during atomic clock operation, the probe field frequency $\omega$ is stabilized to the extreme transitions $|m_{\rm g}=-F\rangle\rightarrow |m_{\rm e}=-F\rangle$ and $|m_{ \rm g}=+F\rangle\rightarrow |m_{\rm e}=+F\rangle$, the presence of two circular components $a^{(\pm 1)}$ will lead to ac Stark shifts of the lower ($|m_{\rm g}=\pm F\rangle$) and upper ($|m_{\rm e}= \pm F\rangle$) Zeeman sublevels due to non-resonant interaction with the neighboring magnetic sublevels $|m_{\rm g}=\pm F\mp 1\rangle$ and $|m_{\rm e}=\pm F\mp 1\rangle$. As a result, instead of Eq.~(\ref{w_FF}) we get the following expressions for the frequencies $\omega^{}_{-F-F}$ and $\omega^{}_{+F+F}$:
\begin{eqnarray}\label{w_FF_ac}
\omega^{}_{-F-F}&=&\omega^{}_0-(\Delta^{}_{\rm e}-\Delta^{}_{\rm g})F+\bar{\delta}^{}_{-F-F}\,,\nonumber\\
\omega^{}_{+F+F}&=&\omega^{}_0+(\Delta^{}_{\rm e}-\Delta^{}_{\rm g})F+\bar{\delta}^{}_{+F+F}\,,
\end{eqnarray}
where $\bar{\delta}^{}_{-F-F}$ and $\bar{\delta}^{}_{+F+F}$ are the corresponding ac Stark shifts:
\begin{eqnarray}\label{D_FF_ac}
\bar{\delta}^{}_{-F-F}&=&\frac{|\Omega^{}_{+}|^2}{4\Delta^{}_{\rm e}}-\frac{|\Omega^{}_{-}|^2}{4\Delta^{}_{\rm g}}\,,\nonumber\\
\bar{\delta}^{}_{+F+F}&=&\frac{|\Omega^{}_{+}|^2}{4\Delta^{}_{\rm g}}-\frac{|\Omega^{}_{-}|^2}{4\Delta^{}_{\rm e}}\,.
\end{eqnarray}
The expressions (\ref{w_FF_ac}) and (\ref{D_FF_ac}) are derived under the condition
\begin{equation}\label{D_eg}
|\Delta^{}_{\rm e,g}|\gg |\Omega^{}_{\pm}|\,,
\end{equation}
which holds well in real experiments and corresponds to a non-resonant interaction for circular components $E a^{(\pm 1)}{\bf e}^{}_{\pm 1}$.

Thus, the clock frequency
\begin{equation}\label{w_clock_ac}
\omega_{\rm clock}=\frac{\omega^{}_{-F-F}+\omega^{}_{+F+F}}{2}=\omega^{}_0+\bar{\delta}^\text{(el-ind)}_{\rm ac}\,,
\end{equation}
becomes shifted relatively to the unperturbed frequency $\omega^{}_0$ by the value
\begin{eqnarray}\label{sh}
\bar{\delta}^\text{(el-ind)}_{\rm ac}&=&\frac{\bar{\delta}^{}_{-F-F}+\bar{\delta}^{}_{+F+F}}{2}=\frac{\Delta^{}_{\rm g}+\Delta^{}_{\rm e}}{8\Delta^{}_{\rm g}\Delta^{}_{\rm e}}(|\Omega^{}_{+}|^2-|\Omega^{}_{-}|^2)= \nonumber \\
&& \frac{|d^{}_{\rm eg}E|^2}{\hbar^2 (F+1)}\frac{\Delta^{}_{\rm g}+\Delta^{}_{\rm e}}{8\Delta^{}_{\rm g}\Delta^{}_{\rm e}}(|a^{(+1)}|^2-|a^{(-1)}|^2)\,,
\end{eqnarray}
which we will denote as the ellipticity-induced shift. The nature of which is an ac Stark shift through interaction with neighboring Zeeman sublevels. Using the expression for $|\Omega^{}_{0}|$ in Eq.~(\ref{W}), the formula (\ref{sh}) can be rewritten as
\begin{equation}\label{sh_2}
\bar{\delta}^\text{(el-ind)}_{\rm ac}=|\Omega^{}_{0}|^2\frac{\Delta^{}_{\rm g}+\Delta^{}_{\rm e}}{8F\Delta^{}_{\rm g}\Delta^{}_{\rm e}}\frac{|a^{(+1)}|^2-|a^{(-1)}|^2}{|a^{(0)}|^2}\,,
\end{equation}
which we will analyze further.

First of all, using vector notations, we represent in an invariant form the following expression
\begin{equation}\label{a}
\frac{|a^{(+1)}|^2-|a^{(-1)}|^2}{|a^{(0)}|^2}=\frac{i([{\bf a}\times{\bf a}^{\ast}]\cdot {\bf n}^{}_{\rm B})}{|{\bf a}\cdot {\bf n}^{}_{\rm B}|^2}\,,
\end{equation}
where $[{\bf a}\times{\bf a}^{\ast}]$ denotes the cross product of two vectors ${\bf a}$ and ${\bf a}^{\ast}$, and ${\bf n}^{}_{\rm B}={\bf B}/|{\bf B}|$ is the unit orientation vector of the magnetic field ${\bf B}$. As shown in Fig.~\ref{Fig2}(b), the degree of ellipticity of the polarization vector ${\bf a}$ can be parametrized by the angular parameter $\varepsilon$
\begin{equation}\label{eps}
{\bf a}=\cos(\varepsilon){\bf e}'_{x}+i\sin(\varepsilon){\bf e}'_{y}\,,
\end{equation}
where ${\bf e}'_{x,y}$ are the Cartesian unit vectors oriented along the axes of the polarization ellipse. In this case we have:
\begin{equation}\label{aa}
i[{\bf a}\times{\bf a}^{\ast}]=\sin (2\varepsilon){\bf n}^{}_{\rm k}\,,
\end{equation}
where ${\bf n}^{}_{\rm k}={\bf e}'_{z}$ is the unit vector orthogonal to the plane of the polarization ellipse, i.e. directed along the wave vector of the probe field. This, in turn, leads to
\begin{equation}\label{aaB}
i([{\bf a}\times{\bf a}^{\ast}]\cdot {\bf n}^{}_{\rm B})=\sin(2\varepsilon)({\bf n}^{}_{\rm k}\cdot {\bf n}^{}_{\rm B})=\sin(2\varepsilon)\sin(\xi)\,,
\end{equation}
where $\xi$ is the angle between the vector ${\bf B}$ and the plane of the polarization ellipse ${\bf a}$ [see Fig.~\ref{Fig2}(a)]. In addition, using Fig.~\ref{Fig2}(b), we obtain an expression for $|{\bf a}\cdot {\bf n}^{}_{\rm B}|^2$
\begin{eqnarray}\label{a0}
&&|{\bf a}\cdot {\bf n}^{}_{\rm B}|^2=|{\bf a}\cdot {\bf n}^{\rm (ell)}_{\rm B}|^2=\nonumber \\
&&\left[\cos^2(\varepsilon)\cos^2(\varphi)+\sin^2(\varepsilon)\sin^2(\varphi)\right]\cos^2(\xi)\,,
\end{eqnarray}
where ${\bf n}^{\rm (ell)}_{\rm B}$ is the projection of the unit vector ${\bf n}^{}_{\rm B}$ onto the plane of the polarization ellipse, and $\varphi$ is the angle between ${\bf n}^{\rm (ell)}_{\rm B}$ and the major axis of the polarization ellipse.

Thus, using (\ref{aaB}) and (\ref{a0}) in Eq.~(\ref{a}), the shift (\ref{sh_2}) can be calculated by the formula
\begin{equation}\label{sh_3}
\bar{\delta}^\text{(el-ind)}_{\rm ac}= |\Omega^{}_{0}|^2\frac{\Delta^{}_{\rm g}+\Delta^{}_{\rm e}}{8F\Delta^{}_{\rm g}\Delta^{}_{\rm e}} \frac{\sin(2\varepsilon)\sin(\xi)}{|{\bf a}\cdot {\bf n}^{}_{\rm B}|^2}\,.
\end{equation}
Based on this expression, one can find three conditions where the shift vanishes, $\bar{\delta}^\text{(el-ind)}_{\rm ac}=0$:\\
1) purely linear polarization, $\varepsilon=0$, for any value of $\xi$;\\
2) $\xi=0$ for any ellipticity $\varepsilon$, when the magnetic field vector ${\bf B}$ lies in the plane of the polarization ellipse;\\
3) $g^{}_{\rm g}=-g^{}_{\rm e}$ (when $\Delta^{}_{\rm g}+\Delta^{}_{\rm e}=0$) for any $\varepsilon$ and $\xi$. However, this ``exotic'' variant does not occur for real atomic transitions $F_{\rm g}=F\rightarrow F_{\rm e}=F$.

\begin{figure}[t]
\centerline{\scalebox{1.35}{\includegraphics{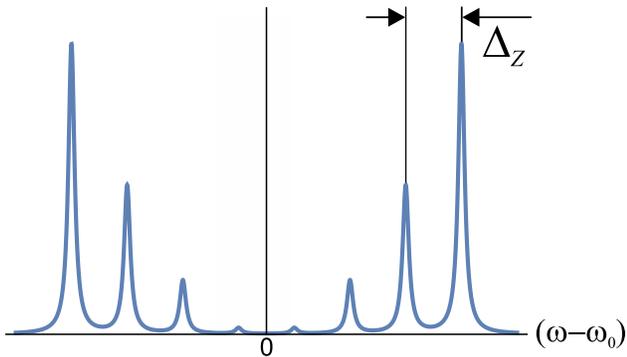}}}\caption{
Illustration of a series of optical resonances $|m_{\rm g}=m\rangle\rightarrow |m_{\rm e}=m\rangle$ ($-F\leq m\leq F$) in a linearly polarized field for $F_{\rm g}=F\rightarrow F_{\rm e}=F$ transitions, using $F=7/2$ as an example. The frequency distance between adjacent resonances is defined as $\Delta^{}_{Z}=|\Delta^{}_{\rm g}-\Delta^{}_{\rm e}|$.} \label{Fig3}
\end{figure}

Regarding to atomic clocks, the residual shift $\bar{\delta}^\text{(el-ind)}_{\rm ac}$ appears due to some uncontrollability of small values $\varepsilon$ and $\xi$. Thus, the signs of the quantities $\varepsilon$, $\xi$, and therefore the sign of $\bar{\delta}^\text{(el-ind)}_{\rm ac}$ should also be considered as uncontrolled. Therefore, we will be interested in the absolute value $|\bar{\delta}^\text{(el-ind)}_{\rm ac}|$. In addition, consider the complete picture of the resonance series over all Zeeman sublevels shown in Fig.~\ref{Fig3}, where the distance between adjacent resonances is equal to $\Delta^{}_{Z}=|\Delta^{}_{\rm g}-\Delta^{}_{\rm e}|$. This allows us to represent the shift in the form
\begin{eqnarray}\label{sh_4}
|\bar{\delta}^\text{(el-ind)}_{\rm ac}|&=&\frac{|\Omega^{}_{0}|^2}{\Delta^{}_{Z}}\frac{|\Delta^{2}_{\rm g}-\Delta^{2}_{\rm e}|}{8F|\Delta^{}_{\rm g}\Delta^{}_{\rm e}|}\frac{|\sin(2\varepsilon)\sin(\xi)|}{|{\bf a}\cdot {\bf n}^{}_{\rm B}|^2}= \nonumber \\
&& \frac{|\Omega^{}_{0}|^2}{\Delta^{}_{Z}}\frac{|g^{2}_{\rm g}-g^{2}_{\rm e}|}{8F|g^{}_{\rm g}g^{}_{\rm e}|}\frac{|\sin(2\varepsilon)\sin(\xi)|}{|{\bf a}\cdot {\bf n}^{}_{\rm B}|^2}\,.
\end{eqnarray}
Typically for atomic clocks, the experimental conditions are close to the ideal case of linear probe field polarization  ${\bf E}$ that is parallel to the magnetic field ${\bf B}$. Therefore, under small non-idealities, the condition of small value for all three angular parameters $|\varepsilon,\xi,\varphi|\ll 1$ can be assumed, which allows us to use the following approximations
 \begin{equation}\label{approx}
\sin(2\varepsilon)\approx 2\varepsilon,\;\; \sin(\xi)\approx \xi,\;\; |{\bf a}\cdot {\bf n}^{}_{\rm B}|^2\approx 1-\varepsilon^2-\varphi^2-\xi^2,
 \end{equation}
which, in turn, leads to the main approximation for the ellipticity-induced shift:
\begin{equation}\label{sh_5}
|\bar{\delta}^\text{(el-ind)}_{\rm ac}|\approx A\frac{|\Omega^{}_{0}|^2}{\Delta^{}_{Z}}|\varepsilon\xi|\,,
\end{equation}
where the coefficient
\begin{equation}\label{A}
A=\frac{|g^{2}_{\rm g}-g^{2}_{\rm e}|}{4F|g^{}_{\rm g}g^{}_{\rm e}|}\,,
\end{equation}
is a characteristics of the specific clock transition, and the angular parameters $\varepsilon$ and $\xi$ in Eq.~(\ref{sh_5}) are defined in radian units. We remind that $|\Omega^{}_{0}|$ is the Rabi frequency for clock transitions $|m_{\rm g}=-F\rangle\rightarrow |m_{\rm e}=-F\rangle$ and $|m_{\rm g}=+F\rangle\rightarrow |m_{\rm e}=+F\rangle$ between the extreme Zeeman sublevels.

\begin{table}[h]
\caption{Values of the coefficient $A$ [see Eq.~(\ref{A})] for clock transitions $F_{\rm g}=F\rightarrow F_{\rm e}=F$ in atoms and ions for which the strongly-forbidden transition $^1$S$_0\rightarrow{}^3$P$_0$ currently used for clock operation. We used the values of $g$-factors from Refs.~\cite{Porsev_2004,Lu_2018,Rosenband_2007,Becker_2001}.} \label{table}
\begin{tabular}{|c|c|c|c|c|c|}
\hline
 & $^{171}$Yb & $^{173}$Yb & $^{87}$Sr & $^{27}$Al$^+$ & $^{115}$In$^+$ \\
\hline
$F$ & 1/2 & 5/2 & 9/2 & 5/2 & 9/2 \\
\hline
$A$ & 0.85 & 0.18 & 0.053 & 0.209 & 0.045 \\
\hline
\end{tabular}
\end{table}

Table~\ref{table} presents the values of the parameter $A$ for the optical clock transitions in some atoms and ions. Note that, as follows from Eq.~(\ref{A}), the sensitivity to the ellipticity-induced shift for $F_{\rm g}=F\rightarrow F_{\rm e}=F$ transitions decreases with increasing of $F$.

Let us now estimate the magnitude of possible values of the angular parameters $\varepsilon$ and $\xi$. Under typical experimental conditions, after the passage of light through vacuum windows, a degree of ellipticity can be estimated as (e.g., see Refs.~ \cite{Nemitz_2019,Mehlstaubler})
\begin{equation}\label{ratio}
\frac{I^{}_{x'}}{I{}_{y'}}=\tan^2(\varepsilon)\approx\varepsilon^2 \sim 0.01\,,\;\Rightarrow\;  |\varepsilon|\sim 0.1~\text{rad}\,,
\end{equation}
using the ratio of the intensities of the orthogonal components $I^{}_{x'}$ and $I{}_{y'}$ along the main axes of the polarization ellipse [see Fig.~\ref{Fig2}(b)].  We also assume that the possible uncertainty of the magnetic field orientation ${\bf n}^{}_B$ with respect to the polarization vector ${\bf a}$ can be several angular degrees, i.e., $|\xi|\sim 0.1$ rad. Thus, we obtain the following general estimate for the angular parameters:
\begin{equation}\label{ef}
|\varepsilon\xi|\sim 0.01\,,
\end{equation}
which we will use in further evaluations.

\section{Total ac Stark shift}

Since the presented ellipticity-induced shift is proportional to the square of the electric field, $\bar{\delta}^\text{(el-ind)}_{\rm ac}\propto |E|^2$, it can be considered as an additional, transition internal ac Stark shift. Simultaneously, there always exists the well-known standard ac Stark shift, $\bar{\delta}^\text{(off-res)}_{\rm ac}\propto |E|^2$, due to the interaction of the probe field with other far-off-resonant atomic levels, and which can be represented as
\begin{equation}\label{ac}
\bar{\delta}^\text{(off-res)}_{\rm ac}=\alpha |\Omega^{}_{0}|^2,
\end{equation}
where $\alpha$ is some proportionality factor. Therefore, to estimate the uncertainty in atomic clocks, one must always consider the total ac Stark shift
\begin{equation}\label{ac_tot}
\bar{\delta}^{\rm (tot)}_{\rm ac}=\bar{\delta}^\text{(el-ind)}_{\rm ac}+\bar{\delta}^\text{(off-res)}_{\rm ac}= |\Omega^{}_{0}|^2 K\,,
\end{equation}
where the parameter $K$ in our case is defined as
\begin{eqnarray}\label{K}
K&=&\frac{\Delta^{}_{\rm g}+\Delta^{}_{\rm e}}{8F\Delta^{}_{\rm g}\Delta^{}_{\rm e}}\frac{\sin(2\varepsilon)\sin(\xi)}{|{\bf a}\cdot {\bf n}^{}_{\rm B}|^2} +\alpha\approx \nonumber \\
&& \frac{\Delta^{}_{\rm g}+\Delta^{}_{\rm e}}{4F\Delta^{}_{\rm g}\Delta^{}_{\rm e}}\,\varepsilon\xi +\alpha\,,
\end{eqnarray}
according to the formula (\ref{sh_3}), as well as the conditions $|\varepsilon,\xi|\ll 1$ and $|{\bf a}\cdot {\bf n}^{}_{\rm B}|^2\approx 1$.

Unlike the ellipticity-induced shift, $\bar{\delta}^\text{(off-res)}_ {\rm ac}$ has a fixed sign. For example, in the case of the clock transition $^1$S$_0$$\to$$^3$P$_0$, the quantity $\bar{\delta}^\text{(off-res)}_{\rm ac}$ depends very weakly on the ellipticity parameter $\varepsilon$. Therefore, the absolute value of the total ac Stark shift can take two values
\begin{equation}\label{ac_mod}
|\bar{\delta}^{\rm (tot)}_{\rm ac}|=||\bar{\delta}^\text{(off-res)}_{\rm ac}|\pm |\bar{\delta}^\text{(el-ind)}_{\rm ac}||\,,
\end{equation}
where the sign ($\pm$) can change depending on the signs of $\varepsilon$ and $\xi$ [see Eq.~(\ref{sh_3})]. However, if we assume that a small degree of ellipticity $\varepsilon$ appears in experiments in an uncontrolled way (as well as $\xi$), then the maximal value should be used for metrological estimates
 \begin{equation}\label{ac_metr}
\bar{\delta}^{\rm (met)}_{\rm ac}=|\bar{\delta}^\text{(off-res)}_{\rm ac}|+|\bar{\delta}^\text{(el-ind)}_{\rm ac}|\,,
\end{equation}
which is more conservative (from a metrological viewpoint). It is also interesting to note that, based on Eq.~(\ref{K}), one can choose such values of $\varepsilon$, $\xi$ and $|{\bf B}|$ for which $K=0$, i.e., the total ac Stark shift vanishes, $\bar{\delta}^{\rm (tot)}_{\rm ac}=0$.

\begin{figure}[t]
\centerline{\scalebox{1.1}{\includegraphics{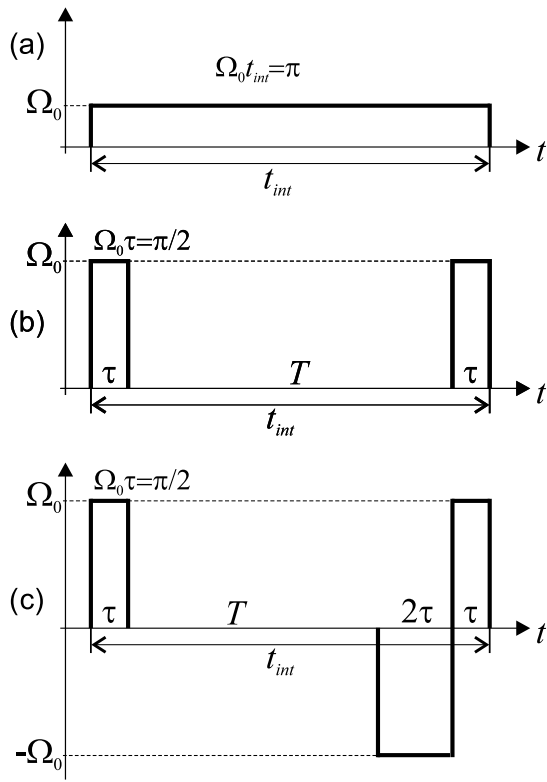}}}\caption{
Scheme of three variants of spectroscopy with the same total interrogation time $t^{}_{int}$:\\
(a) $\pi$-pulse in standard Rabi spectroscopy;\\
(b) the standard Ramsey sequence with two $\pi/2$-pulses, $T=(t^{}_{int}-2\tau)$ is the free evolution time;\\
(c) hyper-Ramsey sequence \cite{yudin2010} including composite pulse, $T=(t^{}_{int}-4\tau)$ is the free evolution time.} \label{Fig4}
\end{figure}

Below we will carry out a comparative analysis of three different spectroscopic schemes (see Fig.~\ref{Fig4}) with the same total interrogation time $t^{}_{int}$.

\section{Rabi spectroscopy}

As applied to Rabi spectroscopy with a single $\pi$-pulse [see Fig.~\ref{Fig4}(a)], the frequency shift of the clock transition $\bar{\delta}^{\rm (Rabi)}_{\rm ac}$ is determined by the shift $\bar{\delta}^{\rm (tot)}_{\rm ac}$ [see Eq.~(\ref{ac_tot})], which can be represented as:
\begin{equation}\label{D_Rabi}
\bar{\delta}^{\rm (Rabi)}_{\rm ac}=\bar{\delta}^{\rm (tot)}_{\rm ac}\approx\frac{\pi^2}{t^2_{int}}K\,,
\end{equation}
taking into account the condition $\Omega^{}_0 t^{}_{int}=\pi$ in Eq.~(\ref{sh_5}).

Analyzing various papers, we found that the frequency shift $|\bar{\delta}^{\rm (Rabi)}_{\rm clock}|$ under certain experimental conditions can reach several mHz (when using Rabi spectroscopy), which noticeably exceeds the fractional level of $10^{-18}$. Thus, this shift must always be estimated, and further, depending on this estimate, should be included/excluded in the budget of uncertainties for modern ultra-precise atomic clocks.

For example, consider the experiment Ref.~\cite{Kobayashi_2020}, where the clock transition $F_{\rm g}=1/2\rightarrow F_{\rm e}=1/2$ in $^{171}$Yb was used
under $t^{}_{int}=40$~ms Rabi $\pi$-pulse and dc magnetic field of 65~$\mu$T. In this case, we have $|\Omega^{}_0|/2\pi =12.5$~Hz and $\Delta^{}_Z/2\pi =260$~Hz. Taking into account the value $|\varepsilon\xi|\sim 0.01$ of Eq. (\ref{ef}), we obtain a possible shift $|\bar{\delta}^\text{(el-ind)}_{\rm ac}|\sim 5$~mHz, which corresponds to the fractional value of $9.6\times 10^{-18}$ for the 578~nm clock transition. At the same time, the standard ac Stark shift $|\bar{\delta}^\text{(off-res)}_{\rm ac}|$ is estimated in Ref.~\cite{Kobayashi_2020} at the level of $4\times 10^{ -18}$. Thus, according to Eq.~(\ref{ac_metr}), the total ac Stark shift $|\bar{\delta}^{\rm (Rabi)}_{\rm ac}|$ can potentially reach the value of $1.4\times10^{-17}$.

\section{Standard Ramsey spectroscopy}

Let us consider standard Ramsey spectroscopy with two identical $\pi/2$ pulses separated by a free evolution time $T$ [see Fig.~\ref{Fig4}(b)], for which the central Ramsey resonance will be used for frequency stabilization in atomic clocks. Therefore, the frequency shifts $\bar{\delta}^{}_{-F-F}$ and $\bar{\delta}^{}_{+F+F}$ due to the ellipticity of the light, as well as the standard ac Stark shift $\bar{\delta}^\text{(off-res)}_{\rm ac}$, will occur only during Ramsey pulses of duration $\tau$, while during the free interval $T$ these shifts are absent. In this case, the shift of the central Ramsey resonance $\bar{\delta}^{\rm (Rams)}_{\rm ac}$ is no longer equal to $\bar{\delta}^{\rm (tot)}_ {\rm ac}$ (as for Rabi spectroscopy), but defined as (see Ref.~\cite{yudin2010,tai09})
\begin{equation}\label{D_Rams}
\bar{\delta}^{\rm (Rams)}_{\rm ac}\approx\frac{2\bar{\delta}^{\rm (tot)}_{\rm ac}}{2+|\Omega^{}_0|T}\,.
\end{equation}
Using the expression (\ref{ac_tot}) and the condition $|\Omega^{}_0|\tau=\pi/2$, the shift (\ref{D_Rams}) can be represented in the form
\begin{equation}\label{D_Rams2}
\bar{\delta}^{\rm (Rams)}_{\rm ac}\approx\frac{\pi^2}{2\tau^2[2+\pi T/(2\tau)]} K\,.
\end{equation}
Comparing Eq.~(\ref{D_Rams2}) and Eq.~(\ref{D_Rabi}), it is easy to show the inequality
\begin{equation}\label{Rams_Rabi}
|\bar{\delta}^{\rm (Rams)}_{\rm ac}|>|\bar{\delta}^{\rm (Rabi)}_{\rm ac}|\,.
\end{equation}
Indeed, taking into account that $T=(t^{}_{int}-2\tau)$, consider the ratio of the quantities (\ref{D_Rams2}) and (\ref{D_Rabi}) for the same interrogation time $t^{}_{int}$:
\begin{equation}\label{ratio}
\frac{\bar{\delta}^{\rm (Rams)}_{\rm ac}}{\bar{\delta}^{\rm (Rabi)}_{\rm ac}}=\frac{1}{2(\tau/t^{}_{int})^2\left[2-\pi+0.5\pi(\tau/t^{}_{int})^{-1}_{}\right]}\,,
\end{equation}
where $\tau <0.5t^{}_{int}$.

\begin{figure}[t]
\centerline{\scalebox{1.15}{\includegraphics{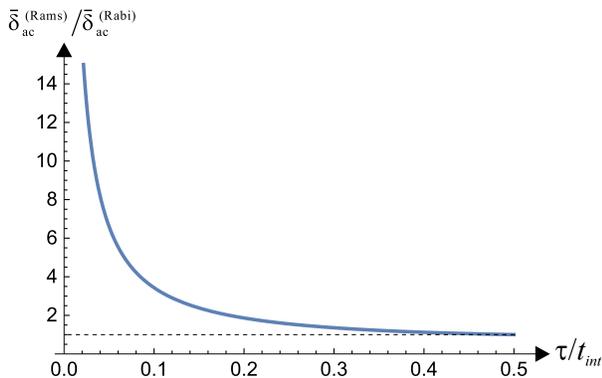}}}\caption{
Dependence of the ratio $\bar{\delta}^{\rm (Rams)}_{\rm ac}/\bar{\delta}^{\rm (Rabi)}_{\rm ac}$ on $\tau/t^{}_{int}$.} \label{Fig5}
\end{figure}

Fig.~\ref{Fig5} shows the dependence of the ratio (\ref{ratio}) on $\tau/t^{}_{int}$, from which the validity of the inequality (\ref{Rams_Rabi}) follows, since $\bar{\delta}^{\rm (Rams)}_{\rm ac}/\bar{\delta}^{\rm (Rabi)}_{\rm ac}>1$. We see that the shift $|\bar{\delta}^{\rm (Rams)}_{\rm ac}|$ increases with decreasing $\tau$. For example, in the case of $\tau <0.1 t^{}_{int}$, $|\bar{\delta}^{\rm (Rams)}_{\rm ac}|>3.4 |\bar{\delta}^{\rm (Rabi)}_{\rm ac}|$. Note that under the condition $\tau\ll t^{}_{int}$, which is typical for Ramsey spectroscopy, the expressions (\ref{D_Rams})-(\ref{D_Rams2}) can be represented as
\begin{equation}\label{D_Rams3}
\bar{\delta}^{\rm (Rams)}_{\rm ac}\approx\frac{2}{T}\frac{\bar{\delta}^{\rm (tot)}_{\rm ac}}{|\Omega^{}_0|}\approx\frac{\pi}{\tau T} K\,.
\end{equation}
Thus, we can assert that using standard Ramsey spectroscopy, any shift (including the ellipticity-induced shift) proportional to the the probe field intensity (i.e. $\propto |E|^2$) will be noticeably larger compared to Rabi spectroscopy. Moreover, since the value of the coefficient $K$ in Eq.~(\ref{ac_tot}) did not play any role in proving, this statement about ac Stark shift is general when comparing Ramsey spectroscopy and Rabi spectroscopy for any one-photon transitions.

In this context, it is interesting to consider the experiment in Ref.~\cite{McGrew_2018}, where the clock transition $F_{\rm g}=1/2\rightarrow F_{\rm e}=1/2$ in $^{171}$Yb was used under $t^{}_{int}=560$~ms Rabi $\pi$-pulse and dc magnetic field of 0.1~mT. In this case, we have $|\Omega^{}_0|/2\pi =0.9$~Hz and $\Delta^{}_Z/2\pi =400$~Hz. Taking into account the value of Eq. (\ref{ef}), we obtain a possible shift of  $|\bar{\delta}^\text{(el-ind)}_{\rm ac}|\sim 0.017$~mHz, which corresponds to the fractional value of $3.3\times 10^{-20}$ for the 578~nm clock transition. At the same time, the standard ac Stark shift $\bar{\delta}^\text{(off-res)}_{\rm ac}$ is estimated in Ref.~\cite{McGrew_2018} as $2\times 10^{-20}$. Therefore, according to Eq.~(\ref{ac_metr}), the total ac Stark shift $\bar{\delta}^{\rm (Rabi)}_{\rm ac}$ is estimated at the fractional level of $5.3\times 10^{-20}$. On the other hand, Ref.~\cite{McGrew_2018} also presents an experiment with $T=510$~ms free-evolution-time Ramsey spectroscopy, when the total interrogation time for Ramsey spectroscopy is $t^{}_{int}=560$~ms, i.e. the same as for Rabi spectroscopy \cite{Ludlow}.
In this case, we have $\tau=25$~ms duration of each Ramsey $\pi/2$ pulse, which gives the ratio $\tau/t^{}_{int}=0.0446$. Then, according to Eq.~(\ref{ratio}), we find the relationship $\bar{\delta}^{\rm (Rams)}_{\rm ac}=7.38\,\bar{\delta}^ {\rm (Rabi)}_{\rm ac}$. As a result, the total ac Stark clock shift $|\bar{\delta}^{\rm (Rams)}_{\rm ac}|$ in Ramsey spectroscopy can potentially reach the fractional value of $ 4\times10^{-19}$, which is no longer negligible for the uncertainty budget in Ref.~\cite{McGrew_2018}.

Note also that when using Rabi spectroscopy or standard Ramsey spectroscopy, the ellipticity-induced shift $\bar{\delta}^\text{(el-ind)}_{\rm ac}$ can be reduced by increasing the magnetic field [i.e., increasing the value of $\Delta^{}_Z$ in the denominator in Eq.~(\ref{sh_5})]. However, in this case the quadratic Zeeman shift and its fluctuations can increase significantly, while the standard ac Stark shift $\bar{\delta}^\text{(off-res)}_{\rm ac}$ will not change.

\section{hyper-Ramsey spectroscopy}

Let us now consider the hyper-Ramsey spectroscopy, which was proposed in Ref.~\cite{yudin2010} and first experimentally implemented in Ref.~\cite{hunt12}. This method consists of using a sequence of pulses, of which one is a composite pulse with inverted phase [see Fig.~\ref{Fig4}(c)]. As applied to our case, the main advantage of this hyper-Ramsey sequence is that the frequencies of the central Ramsey resonances $\omega^{}_{-F-F}$ and $\omega^{}_{+F+F }$ between the extreme Zeeman sublevels will be shifted by values
\begin{eqnarray}\label{h_Rams_FF}
&&\bar{\delta}^\text{(h-Rams)}_{-F-F}\approx\frac{\pi}{T}\left(\frac{\bar{\delta}^{}_{-F-F}+\bar{\delta}^\text{(off-res)}_{\rm ac}}{|\Omega^{}_0|}\right)^3,\nonumber \\ &&\bar{\delta}^\text{(h-Rams)}_{+F+F}\approx\frac{\pi}{T}\left(\frac{\bar{\delta}^{}_{+F+F}+\bar{\delta}^\text{(off-res)}_{\rm ac}}{|\Omega^{}_0|}\right)^3,
\end{eqnarray}
respectively, as follows from Ref.~\cite{yudin2010}. For standard experimental conditions with $F_{\rm g}=F\rightarrow F_{\rm e}=F$ clock transitions in odd isotopes, the following inequality is usually satisfied
\begin{equation}\label{DW_ratio}
\left|\frac{\bar{\delta}^{}_{\pm F\pm F}+\bar{\delta}^\text{(off-res)}_{\rm ac}}{\Omega^{}_0}\right|\sim \left|\frac{\bar{\delta}^{\rm (tot)}_{\rm ac}}{\Omega^{}_0}\right|<0.001\,.
\end{equation}
Therefore, due to the cubic dependence in Eq.~(\ref{h_Rams_FF}) on small values [see Eq.~(\ref{DW_ratio})], the ac Stark shift for the hyper-Ramsey scheme
\begin{equation}\label{h-Rams_clock}
\bar{\delta}^\text{(h-Rams)}_{\rm ac}=\frac{\bar{\delta}^\text{(h-Rams)}_{-F-F}+\bar{\delta}^\text{(h-Rams)}_{+F+F}}{2}\,,
\end{equation}
becomes significantly lower than $10^{-19}$ relative to the clock frequency $\omega^{}_0$.

Thus, the use of hyper-Ramsey spectroscopy Ref.~\cite{yudin2010} in ultra-precise atomic clocks is a simple and effective method to radically solve the problem of any ac Stark shift (including the ellipticity-induced shift $\bar{\delta}^\text{(el-ind)}_{\rm ac}$ as well as the standard ac Stark shift $\bar{\delta}^\text{(off-res)}_{\rm ac}$) without changing other experimental conditions and for arbitrary clock transition.

\section*{CONCLUSION}

We considered a systematic shift in atomic optical clocks due to both some uncontrolled ellipticity of the probe field and some uncertainty in the direction of the dc magnetic field vector (ellipticity-induced shift). Using the example of the clock transitions $F_{\rm g}=F\rightarrow F_{\rm e}=F$, it is shown that, in the presence of an uncontrolled ellipticity of the probe field and at a few degrees misalignment of the dc magnetic field, this shift can reach a fractional level in the range of $10^ {-18}$-$10^{-19}$ for Rabi spectroscopy or standard Ramsey spectroscopy. Therefore, in case the alignment of the dc magnetic field and probe field polarisation is not well known, it needs to be taken into account in the uncertainty budget for modern ultra-precise atomic clocks.

In addition, it is shown that when using hyper-Ramsey spectroscopy Ref.~\cite{yudin2010}, the total ac Stark shift (including the ellipticity-induced shift) can be suppressed to a level significantly lower than $10^{-19}$. This is of particular importance for mobile devices, in which the interrogation time of atoms is relatively short, which requires the use of a higher probe field intensity. Also, in mobile devices, a high level of control of the probe field ellipticity and the magnetic field orientation is difficult, that leads to an increase in the value of $|\varepsilon\xi|$ in Eq.~(\ref{sh_5}) and, therefore, to an increase of ellipticity-induced shift $\bar {\delta}^\text{(el-ind)}_{\rm ac}$. Moreover, hyper-Ramsey spectroscopy allows to reduce the dc magnetic field in order to noticeably reduce the quadratic Zeeman shift and its fluctuations. All of these combined can improve the accuracy and long-term stability of mobile devices (compared to using Rabi spectroscopy or standard Ramsey spectroscopy).

Our approach can also be adapted to other cases. For example, in Appendix~A, we considered the clock transitions of the type $F_{\rm g}=F\rightarrow F_{\rm e}=F+1$. Further variants can be considered in a similar way, when for an arbitrary transition $F_{\rm g}\rightarrow F_{\rm e}$ not only extreme, but also intermediate Zeeman sublevels are used. In this case, the polarization of the probe field can be either linear or circular (using two opposite circular polarizations). For the case of circularly polarized probe fields, one needs to consider a possible ac Stark shift due to some non-ideal circular polarization and some deviation of the magnetic field orientation from the wave vector direction of the probe field.

\begin{acknowledgments}
We thank P. Schmidt, C. Lisdat and U. Sterr for useful discussions.

\begin{figure}[t]
\centerline{\scalebox{0.35}{\includegraphics{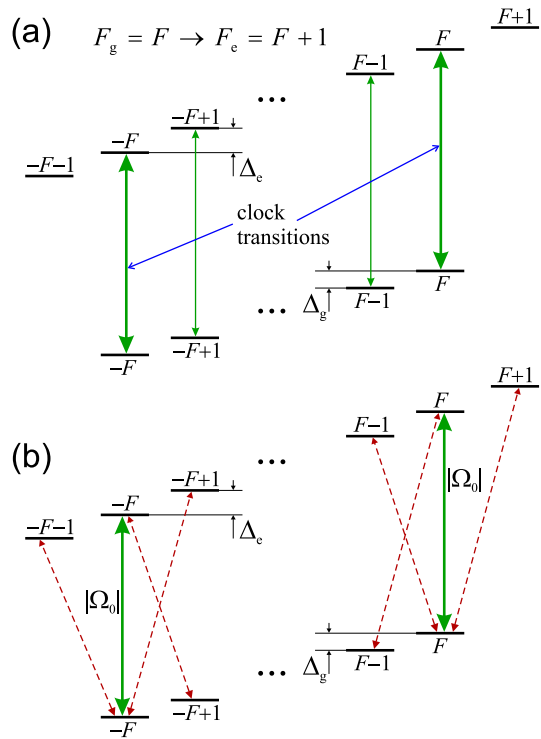}}}\caption{
Scheme of light-induced transitions over Zeeman structure for an optical transition $F_{\rm g}=F\rightarrow F_{\rm e}=F+1$:\\
(a) the ideal case of linear polarization of the probe field ${\bf E}$ directed along the magnetic field ${\bf B}$ (i.e., ${\bf E}||{\bf B}$);\\
(b) the general case of elliptical polarization of the probe field ${\bf E}$ under arbitrary orientation of the magnetic field ${\bf B}$, where the dotted lines mark the light-induced transitions leading to ac Stark shifts (\ref{D_FF_ac}) for the clock transitions $|m_{\rm g}= -F\rangle\rightarrow |m_{\rm e}=-F\rangle$ and $|m_{\rm g}=+F\rangle\rightarrow |m_{\rm e}=+F\rangle$.} \label{Fig6}
\end{figure}

\end{acknowledgments}

\appendix

\section{}
In addition to transitions of the type $F_{\rm g}=F\rightarrow F_{\rm e}=F$, there can, in principle, be the case of the transitions $F_{\rm g}=F\rightarrow F_{\rm e} =F+1$, for which the clock frequency is also determined from the transitions between the extreme Zeeman sublevels $|m_{\rm g}=-F\rangle\rightarrow |m_{\rm e}=-F\rangle$ and $|m_ {\rm g}=+F\rangle\rightarrow |m_{\rm e}=+F\rangle$ in a linearly polarized field [see Fig.~\ref{Fig6}(a)]. This ideal picture of interaction becomes disturbed due to the presence of some uncontrolled ellipticity of the probe field, and also due to some uncertainty of the direction of the magnetic field vector ${\bf B}$ [see Fig.~\ref{Fig6}(b)]. In this case, a residual shift $\bar{\delta}^\text{(el-ind)}_{\rm ac}$ appears, which is also formally described by the expression (\ref{sh_5}). However, the coefficient $A$ is now defined by
\begin{equation}\label{A2}
A=|g^{}_{\rm g}-g^{}_{\rm e}|\left|\frac{F}{4g^{}_{\rm g}}+\frac{2F^2+3F+2}{4(2F+1)g^{}_{\rm e}}\right|\,.
\end{equation}
As can be seen, for the $F_{\rm g}=F\rightarrow F_{\rm e}=F+1$ transitions, the sensitivity to the considered shift increases with $F$ [in contrast to the transitions $F_{\rm g}= F\rightarrow F_{\rm e}=F$, see Eq.~(\ref{A})].


\end{document}